# In-plane Magnetic Anisotropy and Large topological Hall Effect in Self-Intercalated Ferromagnet $Cr_{1.61}Te_2$

Yalei Huang, Na Zuo, Zheyi Zhang, Xiangzhuo Xing[*], Xinyu Yao, Anlei Zhang, Haowei Ma, Chunqiang Xu, Wenhe Jiao, Wei Zhou[*], Raman Sankar, Dong Qian, and Xiaofeng Xu[*]

Yalei Huang, Zheyi Zhang, Wenhe Jiao, Xiaofeng Xu
School of Physics, Zhejiang University of Technology, Hangzhou 310023, China

Na Zuo, Xiangzhuo Xing
Laboratory of High Pressure Physics and Material Science (HPPMS), School of Physics and Physical Engineering, Qufu Normal University, Qufu 273165, China

Xinyu Yao
Materials Genome Institute, Shanghai University, Shanghai 200444, China

Anlei Zhang
College of Science, Nanjing University of Posts and Telecommunications, Nanjing 210023, China

Haowei Ma
College of Mechanical Engineering, Zhejiang University of Technology, Hangzhou 310014, China

Chunqiang Xu
School of Physical Science and Technology, Ningbo University, Ningbo 315211, China

Wei Zhou
School of Electronic and Information Engineering, Changshu Institute of Technology, Changshu 215500, China

Raman Sankar
Institute of Physics, Academia Sinica, Taipei 10617, Taiwan

Dong Qian
Key Laboratory of Artificial Structures and Quantum Control (Ministry of Education), Shenyang National Laboratory for Materials Science, School of Physics and Astronomy, Shanghai Jiao Tong University, Shanghai 200240, China
Tsung-Dao Lee Institute, Shanghai Jiao Tong University, Shanghai 200240, China

E-mail: xuxiaofeng@zjut.edu.cn, xzxing@qfnu.edu.cn and wei.zhou@cslg.edu.cn

Self-intercalated chromium tellurides $Cr_{1+x}Te_2$ have garnered growing attention due to their high-temperature ferromagnetism, tunable spin structures and air stability, all of which are vital for versatile applications in next-generation memory and information technology. Here, we report strong magnetic anisotropy and a large topological Hall effect (THE) in self-intercalated $Cr_{1.61}Te_2$ single crystals, which are both highly desirable properties for future spintronic applications. Our results demonstrate that $Cr_{1.61}Te_2$ is a soft ferromagnet with strong in-plane magnetic anisotropy. Remarkably, distinct THE behaviors are observed in different temperature regimes, reflecting the intricate spin structures and competing exchange interactions. More interestingly, a large topological Hall resistivity, induced by microscopic non-coplanar spin structures, emerges in the temperature range 70-240 K, reaching a maximum value of 0.93 μΩ cm at 150 K. Moreover, a sign-reversed and weak THE is observed at low temperatures below ~70 K, indicating the emergence of an additional topological spin structure with opposite topological charges. This work not only offers valuable insights into the correlation between magnetocrystalline anisotropy and topological phenomena in $Cr_{1+x}Te_2$ systems, but also provides a robust platform for engineering the evolution of complex spin textures that can be leveraged in diverse spintronic device applications.

## 1. Introduction

Van der Waals (vdW) materials have attracted an upsurge of interest in recent years, not only because they host a plethora of intertwined electronic phases with fundamental research value, but they also hold great promise for being integrated into advanced electronic and spintronic devices [1-3]. In particular, vdW ferromagnets, such as $Fe_nGeTe_2$ ($n$ = 3,4,5) [4-6] and $Fe_3GaTe_2$ [7, 8], have attracted great attention and have been widely explored, since their magnetic order can be preserved even when thinned to atomic thickness, contrary to the Mermin-Wagner theorem that predicts long-range ferromagnetic order should be suppressed in the two-dimensional (2D) limit by thermal fluctuations. This feature is exceedingly appealing for low-power, high-performance spintronic applications, including non-volatile memory and logic devices [9, 10].

Nevertheless, many 2D vdW ferromagnets suffer from intrinsic disadvantages such as chemical instability in ambient conditions and a significant reduction of the Curie temperature ($T_C$) when exfoliated, which impedes practical applications[11, 12].

More recently, a class of non-van der Waals materials, namely covalent 2D magnets, has emerged as a promising alternative to 2D vdW ferromagnets[13-15]. Unlike traditional vdW materials, covalent 2D magnets are constructed from vdW layers interleaved with self-intercalated atomic layers, providing additional degrees of freedom for engineering their physical properties and thereby offering a novel approach for the design of advanced spintronic devices. Among these systems, self-intercalated $Cr_{1+x}Te_2$ systems emerge as leading candidates for realizing covalent 2D ferromagnetism in the bulk form[16-18]. As shown in Figure 1a, $Cr_{1+x}Te_2$ systems consist of alternately stacked $CrTe_2$ layers and self-intercalated Cr layers along the *a*-axis or *c*-axis[19, 20]. The parent compound $CrTe_2$ has been reported to exhibit exceptional magnetic properties[21-23]. For example, $1T$-$CrTe_2$ grown by chemical vapor deposition shows an increase in $T_C$ with decreasing thickness, accompanied by a transition of the magnetic anisotropy from an in-plane to out-of-plane orientation[21]. Moreover, monolayer 1T-$CrTe_2$ grown by molecular beam epitaxy exhibits intrinsic ferromagnetism with a $T_C$ as high as 300 K[22]. These advancements make $CrTe_2$ an appealing platform for scalable spintronic device applications. With further incorporation of self-intercalated Cr atoms, the $Cr_{1+x}Te_2$ system exhibits tunable transport properties and magnetic properties that can be implemented in information technology[24-29]. Indeed, the concentration of Cr can effectively modulate the $T_C$ and the magnetocrystalline anisotropy[26], both of which are essential for the practical implementation of spintronic devices. Moreover, the emergence of the THE and topological spin structures, including magnetic skyrmions and non-coplanar spin configurations, in $Cr_{1+x}Te_2$ systems underscores their potential as promising platforms for topological spintronic applications[26, 30-32].

Notably, $Cr_{1+x}Te_2$ crystals with only marginal variations in Cr content exhibit distinct magnetocrystalline anisotropies and markedly different THE behaviors[33, 34]. This versatility underscores the demand for a deeper investigation into the intricate interplay between the magnetocrystalline anisotropy, topological spin structures, and intrinsic transport properties in this system. Such investigations are crucial for both applications and fundamental interest. In this work, we present a comprehensive investigation of the magnetic and transport properties of high-quality $Cr_{1.61}Te_2$ single crystals. Our study demonstrates that $Cr_{1.61}Te_2$ exhibits ferromagnetic order below $T_C \sim$ 326 K, with the magnetic easy axis within the *ab* plane, showing strong magnetocrystalline anisotropy. Detailed transport measurements reveal the presence of a large THE that exhibits distinct behaviors across different temperature regimes, indicative of the evolution of topological spin structures in $Cr_{1.61}Te_2$. Our study demonstrates that $Cr_{1+x}Te_2$ could serve as an improved platform for investigating the evolution of magnetic anisotropy and spin structure that can be harnessed in prospective spin-based magnetic information storage and processing devices.

## 2. Results and Discussion

The single crystals studied in this work were synthesized by using the chemical vapor transport (CVT) method (details given in the Experimental Section). As shown in Figure 1b, the energy-dispersive x-ray spectroscopy (EDX) measurement indicates the actual chemical composition as $Cr_{1.61}Te_2$. The EDX elemental mapping presented in Figure 1c demonstrates a homogeneous distribution of Cr and Te, confirming the absence of elemental segregation during the crystal synthesis. Figure 1d displays the x-ray diffraction (XRD) patterns collected from the as-grown facet of $Cr_{1.61}Te_2$ single crystals, which shows the diffraction peaks from (001) family planes solely. This observation demonstrates that the exposed surface corresponds to the crystalline *ab* plane, indicating the quasi-2D growth mode during the synthesis[16]. The calculated *c*-axis lattice constant is 12.48 Å, which is marginally larger than that of $Cr_3Te_4$ ($x = 0.5$)[35]. This suggests that more Cr atoms are intercalated into the adjoining $CrTe_2$ layers, resulting in an increase of lattice parameters. To further validate the crystal structure of the samples, high-angle annular dark-field scanning transmission electron microscopy (HAADF-STEM) was performed along different crystallographic directions, as shown in Figure 1e and 1f, respectively. Since the HAADF-STEM image intensity is approximately proportional to the average atomic number of the atomic column along the viewing directions, the bright sites correspond to the positions of Te atom, whereas the slightly dimmer sites indicate the locations of Cr atom. The HAADF-STEM image along the [110] direction (Figure 1f) reveals clear evidence of self-intercalated Cr atoms located between the $CrTe_2$ layers. It is observed that the structural model aligns perfectly with the HAADF-STEM images. Collectively, these results confirm the good crystallinity of the samples used in this study.

To investigate the intrinsic magnetic properties of $Cr_{1.61}Te_2$ single crystals, temperature-dependent magnetization (*M-T*) measurements were performed under a magnetic field of 100 Oe, as depicted in Figure 2a. The measurements were conducted in both out-of-plane ($\mu_0H//c$) and in-plane ($\mu_0H//ab$) field configurations, utilizing both zero-field-cooled (ZFC) and field-cooled (FC) modes. It is evident that a ferromagnetic (FM) transition occurs in both field configurations around ~ 320 K. Figure 2b shows the derivative of the *M-T* curve under the out-of-plane field in the FC mode, from which the $T_C$ is determined as ~ 326 K. It is observed that the ZFC curves diverge from the FC curves below $T_C$, suggesting strong magnetocrystalline anisotropy or the presence of spin canting[25, 36]. Moreover, it is noted that, as the temperature decreases, the *M-T* curves exhibit a sharp downturn around $T^{\#}$ ~ 75 K. Similar behaviors were also reported in other $Cr_{1+x}Te_2$ systems[25, 32, 35, 37-40] (Table S1 and Figure S1 in Supporting Information), which is associated with a weak antiferromagnetic (AFM) component superimposed on the strong ferromagnetism. Furthermore, as depicted in Figure 2c and 2d, isothermal magnetization (*M-H*) curves were measured under in-plane and out-of-plane magnetic fields, respectively. In this paper, the effect of a demagnetizing field has been estimated for $\mu_0H_{int} = \mu_0H - 4\pi NM_v$, where $M_v$ is the magnetization per unit volume and the prefactor $4\pi$ is necessary for cgs units[41]. The demagnetization factors are estimated to be $N_{ab}$ = 0.17 ($\mu_0H//ab$) and $N_c$ = 0.77 ($\mu_0H//c$) from the sample shape. Over the entire temperature range (2 to 300 K), the in-plane *M-H* curves consistently exhibit lower saturation fields ($\mu_0H_s$), indicating the presence of in-plane magnetic anisotropy in $Cr_{1.61}Te_2$. Additionally, a negligible coercivity suggests that $Cr_{1.61}Te_2$ is a soft ferromagnet (Figure S2 in Supporting Information). To further explore the

magnetic anisotropy, the magnetic anisotropy energy density $K_U$ was calculated by the expression[42, 43]:

$$K_U = \mu_0 \int_0^{M_s} [H_{ab}(M) - H_c(M)]dM$$

Here, $M_s$ is the saturation magnetization, $H_{ab}$ and $H_c$ represent in-plane and out-of-plane fields, respectively (see the inset of Figure 2e). The temperature-dependent $K_U$ extracted in this manner is summarized in Figure 2e. One can note that the temperature dependence of $K_U$ is non-monotonic over the temperatures measured. For temperatures above $T^\#$, $K_U$ increases with decreasing temperature and reaches a maximum of $3.73 \times 10^5$ J/m$^3$ near $T^\#$. Upon further cooling, $K_U$ decreases slightly, reaching approximately $3.45 \times 10^5$ J/m$^3$ at 2 K. It is noteworthy that even at room temperature, $K_U$ remains as high as $1.15 \times 10^5$ J/m$^3$. These results indicate Cr$_{1.61}$Te$_2$ single crystals exhibit significant magnetocrystalline anisotropy up to room temperature, a property that is important for practical applications. Moreover, Figure 2e suggests that the magnetic domain structures in Cr$_{1.61}$Te$_2$ vary significantly over a wide temperature range of 2-300 K due to the strong influence of magnetic anisotropy[44]. In this light, Cr$_{1.61}$Te$_2$ may serve as an excellent platform for investigating the evolution of spin structure over a broad temperature range.

We further fabricated a Hall device as shown in Figure 3a and conducted magneto-transport measurements. Figure 3b presents the temperature-dependent in-plane resistivity ($\rho$-$T$) of Cr$_{1.61}$Te$_2$, which shows a metallic behavior with the residual resistivity ratio of ~ 5. A clear anomaly sets in around ~320 K in the $\rho$-$T$ curve, which corresponds well to the $T_C$ determined from the magnetization measurement. As the temperature decreases, a kink is also observed, which corresponds to the temperature of $T^\#$ in the $M$-$T$ curves. These features are clearly resolved in the $d\rho/dT$ curve, as illustrated in Figure 3c. Moreover, it is observed that resistivity ($\rho$) follows a quadratic ($T^2$) temperature dependence in the low-temperature regime, implying the dominant electron-electron scattering at low temperatures in Cr$_{1.61}$Te$_2$. This observation conforms to the theoretical studies on weakly itinerant ferromagnetic metals, in which the renormalized spin fluctuations predict that $\rho$ exhibits a $T^2$ dependence on temperature in an itinerant ferromagnet[45].

Figure 3d illustrates the field dependence of magnetoresistance (MR) with the current oriented perpendicular to the magnetic field ($\mu_0H \perp I//ab$). It is noted that all MR curves exhibit a linear profile in the high-field region, denoted by red dashed lines in Figure 3d. The linear negative magnetoresistance (LNMR) is attributed to the linear suppression of spin-magnon scattering under the applied magnetic field[46, 47]. Therefore, the slope of the LNMR can, to some extent, reflect the intensity of the spin-magnon scattering. Figure 3e shows the normalized slope $\alpha(T)/\alpha(300\text{ K})$ of the LNMR over the range of 2-300 K. Generically, the slope of the LMMR increases with increasing temperature in conventional ferromagnets[46]. However, it is observed that the normalized slope displays three distinct regions as temperature decreases. Specifically, for 300 K $>T> T^* \sim$ 240 K, the normalized slope increases with increasing temperature, which is consistent with the theoretical expectation for electron-magnon scattering and the spin-wave damping in conventional ferromagnets. In the intermediate range $T^\# < T < T^*$, the normalized slope becomes nearly temperature-independent, suggesting a possible change in the spin structure that may modify the magnon spectrum and lead to an almost constant electron–magnon scattering within this temperature window. When $T < T^\#$, the normalized slope abruptly decreases in magnitude, likely due to the emergence of a weak AFM evidenced from the magnetization measurement in Figure 2a, which suppresses the magnon generation. Therefore, the normalized slope shown in Figure 3e serves as a sensitive probe of electron–magnon scattering strength, which in turn reflects the underlying evolution of the spin structure in $Cr_{1.61}Te_2$.

Interestingly, as shown in Figure 3d, when $T < T^*$, the MR curves are seen to first increase and then decrease with field, resulting in a peak at $\mu_0H_1$. The extracted $\mu_0H_1$ and the MR values at 7 T are plotted in Figure 3f to illustrate their temperature dependence. With decreasing temperature, $\mu_0H_1$ gradually increases, reaches a maximum at $T^\#$, and then slightly decreases. Meanwhile, the MR value at 7 T also exhibits an extremum at $T^\#$. Generally, the peak-like feature observed in the MR curves of numerous magnetic materials is attributed to the anisotropic magnetoresistance (AMR) effect[48-50]. In ferromagnets with in-plane magnetic anisotropy, when an external magnetic field is applied along the hard axis (*c*-axis), spins are canted away from the easy *ab*-plane. This canting alters the relative orientation between the current and the magnetization from parallel in the absence of field to perpendicular beyond the saturation field. This spin reorientation is accompanied by an increase in resistivity, resulting in a positive MR. However, the $\mu_0 H_1$ in the MR curves differs significantly from the saturation field in the *M-H* curves, as shown in Figure 3g and Figure S3 in Supporting Information. These findings suggest that the peak-like feature in the MR curves of $Cr_{1.61}Te_2$ cannot solely be attributed to the AMR effect. In fact, certain spin structures could also produce a similar peak-like feature in the MR curves[44, 51, 52]. As elaborated below, THE reveals that $Cr_{1.61}Te_2$ hosts rich topological spin structures, which can influence the electron scattering processes and give rise to anomalous MR behaviors. Therefore, we conclude that the peak-like feature observed in the MR curves of $Cr_{1.61}Te_2$ likely originates from a combined effect of AMR effect and the evolution of the topological spin structures.

In magnetic systems hosting topological spin structures, electrons acquire an additional real-space Berry phase when they traverse these structures, resulting in a THE[53-55]. Notably, the topological Hall effect is widely regarded as a key transport signature for the existence of such topological spin structures. To identify the possible topological spin structure in $Cr_{1.61}Te_2$, we measured the magnetic field dependence of the Hall resistivity ($\rho_{yx}$- *H*) at various temperatures, as shown in Figure 4a. It is seen that all $\rho_{yx}$- *H* curves exhibit a nonlinear profile, suggesting the presence of the anomalous Hall effect (AHE). In particular, the AHE exhibits a sign reversal as the temperature decreases below $T^*$, which may be attributed to complex factors such as changes in the band structures and/or spin-dependent scattering[56]. Most remarkably, as seen from Figure 4a and better visualized in Figure 4b for 150 K, a hump is observed in $\rho_{yx}$-*H* curves below $T^*$. The emergent hump features in $\rho_{yx}$-*H* curves suggest an additional contribution to the Hall signal besides the normal Hall effect and the AHE, that is, the THE response. With the inclusion of this THE, the total Hall resistivity ($\rho_{yx}$) in a magnetic system is typically comprised of three components[51, 53, 57],

$$\rho_{yx} = \rho_{yx}^N + \rho_{yx}^A + \rho_{yx}^T = R_0 \mu_0 H + R_s M + \rho_{yx}^T$$

Here, $\rho_{yx}^N$, $\rho_{yx}^A$, and $\rho_{yx}^T$ represent the contributions of the normal Hall effect, AHE,

and THE to the total Hall resistivity, respectively. $R_0$ is the normal Hall coefficient, $R_s$ is the anomalous Hall coefficient. Since topological spin structures and thus THE would vanish in the uniform alignment of magnetic moments when $\mu_0 H > \mu_0 H_s$, we can determine $R_0$ and $R_s$ as the slope and the intercept of the curve $\rho_{yx}/\mu_0 H$ vs $M/\mu_0 H$ above $\mu_0 H_s$. Based on the fitting results above $\mu_0 H_s$, the $\rho_{yx}^N = R_0 \mu_0 H$ and the $\rho_{yx}^A = R_s M$ can be extracted. Consequently, the magnetic field dependence of the topological Hall resistivity ($\rho_{yx}^T$- $H$) curve can be obtained by subtracting the fitted $\rho_{yx}^N + \rho_{yx}^A$ from the experimentally measured total Hall resistivity $\rho_{yx}$, as demonstrated in Figure 4b and Figure S4 in Supporting Information. In addition, as shown in Figure 4c and Figure S5 in Supporting Information, the linear relationship between the anomalous Hall conductivity $\sigma_{yx}^A$ ($= \rho_{yx}^A/[(\rho_{yx}^A)^2 + (\rho_{xx})^2]$) and longitudinal conductivity $\sigma_{xx}(=\rho_{xx}/[(\rho_{yx}^A)^2 + (\rho_{xx})^2])$ below $T^\#$, suggests that the extrinsic skew-scattering mechanism is the primary source for the AHE in this material [58, 59], consistent with the other $Cr_{1+x}Te_2$ systems [18, 32, 34].

The fitting parameters $R_0$ and $R_s$ at various temperatures are summarized in Figure 4d. It can be seen that within the temperature range of 2-300 K, $R_0$ remains positive, indicative of the majority hole carriers. Moreover, when $T < T^*$, $R_s$ changes sign from negative to positive, corresponding to the reversal of the AHE signal. All $\rho_{yx}^T$- $H$ curves in the temperature range of 2-300 K are summarized as a color map in Figure 4e. In this contour plot, the THE in $Cr_{1.61}Te_2$ exhibits three distinct behaviors: when $T > T^*$ (region I), $\rho_{yx}^T$ is nearly zero, whereas for $T^* > T > T^\#$ (region II), $\rho_{yx}^T$ appears in both positive and negative fields with opposite signs. Moreover, when $T < T^\#$ (region III), $\rho_{yx}^T$ decreases in magnitude and reverses its sign. Notably, these three temperature ranges align perfectly with the magnetization and MR measurements. For clarity, the maximum values of $\rho_{yx}^T$ at each temperature ($\rho_{yx}^{T,max}$) are extracted and summarized in Figure 4f. Remarkably, the temperature dependencies of $R_0$, $R_s$, and $\rho_{yx}^A$ all indicate three distinct regions, consistent with the temperature dependence of $\rho_{yx}^{T,max}$. The distinct THE responses suggest that there are three distinct spin structures within the measured temperature range in $Cr_{1.61}Te_2$.

As shown in Figure 4e and f, no THE is detected in region I, demonstrating the absence of topological spin structures in $Cr_{1.61}Te_2$ above $T^*$. It is noteworthy that spin fluctuations and thermal effects may contribute to the suppression of topological spin structures as the temperature approaches $T_C$. Furthermore, as an itinerant ferromagnet, local spins and itinerant electrons in $Cr_{1.61}Te_2$ are strongly coupled; thus, the abrupt reduction in carrier density above $T^*$ (see Figure 4d) could change both electronic and magnetic structures, impeding the formation of topological spin structures. Previous studies have reported that only conventional ferromagnetic domains were identified in $Cr_{1.60}Te_2$ and $Cr_{1.64}Te_2$ at room temperature[26, 27]. Our results seem to align with these observations.

Upon cooling to region II, a large THE is observed in $Cr_{1.61}Te_2$ and reaches a maximum value of $\rho_{yx}^{T,max} = 0.93$ μΩ cm at 150 K. Notably, in this region, the maximum value of $\rho_{yx}^{T,max}$ is comparable to that of $\rho_{yx}^A$ (Figure 4f). This contrasts sharply with observations in most ferromagnetic systems hosting skyrmions or merons, where the AHE typically dominates over the THE[26, 51, 53, 60-63]. This suggests that, unlike in some $Cr_{1+x}Te_2$ systems[26], THE in region II is less likely to be induced by skyrmions. Beyond the *mesoscopic* topological spin structures such as skyrmions and merons, when electrons travel through *microscopic* non-coplanar spin structures in a lattice with finite scalar spin chirality, an emergent magnetic field associated with Berry curvature can induce THE[64]. Generally, the THE arising from *microscopic* non-coplanar spin structures tends to be stronger and comparable in magnitude to the AHE. In contrast, the THE from *mesoscopic* textures like skyrmions is typically much weaker (Figure 5) [32, 34, 65-68]. To explore the origin of THE in $Cr_{1.61}Te_2$, we have performed Lorentz transmission electron microscopy (LTEM) on the [001]-oriented lamella (~100 nm thick) at 120 K. As shown in Figure S6 in Supporting Information, magnetic domains with in-plane magnetization are clearly observed, consistent with our magnetization measurements confirming strong in-plane magnetic anisotropy. Remarkably, no skyrmion-like features were observed under applied magnetic fields, which is conceivable because in-plane magnetic anisotropy is generally unfavorable for stabilizing skyrmions[26]. These results suggest that the large THE observed in $Cr_{1.61}Te_2$ is more likely to originate from *microscopic* non-coplanar spin structures, rather than skyrmions in Region II. Moreover, in Region II, applying a very small out-of-plane magnetic field is sufficient to induce the THE. These findings suggest that, due to the in-plane magnetic anisotropy of $Cr_{1.61}Te_2$, the Cr spins tilt from the *ab*-plane towards the *c*-axis when an out-of-plane magnetic field is applied ($\mu_0 H < \mu_0 H_s$), forming the non-coplanar spin structures in the lattice that induce a significant THE comparable to the AHE (Figure 4g). When $\mu_0 H > \mu_0 H_s$, the collinear spin state is achieved, leading to the vanishing of THE (Figure 4g).

As the temperature further decreases into Region III, a sign reversal of THE is observed, indicating an opposite topological charge compared with Region II. For conduction electrons, the opposite topological charge acts as an opposite fictitious magnetic field, leading to an opposite THE signal. As mentioned above, in general, THE induced by microscopic non-coplanar spin structures is larger and comparable to AHE, while THE induced by mesoscopic non-coplanar spin structures like skyrmions and merons is relatively weak. Notably, the $\rho_{yx}^{T,max}$ significantly decreases in Region III, accounting for merely a few percent (1-3%) of the $\rho_{yx}^{A}$. These results indicate the presence of emergent spin structures in $Cr_{1.61}Te_2$ below $T^{\#}$. These results suggest that THE in this region is unlikely to be induced by microscopic non-coplanar spin structures. Moreover, as shown in Figure S7 in Supporting Information, it is known that skyrmions are likely to be stabilized with out-of-plane magnetic anisotropy in a centrosymmetric system, while $Cr_{1.61}Te_2$ exhibits in-plane magnetic anisotropy. In comparison, periodic circular and cross Bloch lines with a fixed equilibrium distance can be spontaneously generated in the domain walls at zero magnetic field in single-crystal ferromagnets with the in-plane magnetic anisotropy, namely merons and antimerons. For instance, merons and antimeron have been observed in $Fe_5GeTe_2$ with in-plane magnetic anisotropy[69, 70]. Thus, we speculate that merons and antimerons could be a possible source of the low-temperature THE in region III. Alternatively, the emergence of a weak AFM component below 75 K, evidenced from Figure 2a, may modify the magnetic ground state and influence the local scalar spin chirality, resulting in a reversed emergent magnetic field and thus the sign-reversed THE. This effect differs from the typical skyrmion/meron scenario and points toward a complex non-coplanar spin texture modulated by the AFM–FM coexistence. This conjecture merits further studies in the future. Due to experimental restrictions, our LTEM measurements cannot be performed below 88 K, impeding the direct imaging of low- temperature spin structures.

To contextualize the differences and similarities among various $Cr_{1+x}Te_2$ systems reported in the literature and in our present study[26, 32-34, 71], we have summarized the key parameters — including Cr content, temperature regimes, magnetic anisotropy, maximum THE magnitude, and the proposed spin structures—in the Table 1. As seen, we note that out-of-plane magnetic anisotropy promotes the formation of skyrmions, whereas in-plane magnetic anisotropy favors the development of microscopic non-coplanar spin structures rather than skyrmions. This comparison further highlights that the concentration of self-intercalated Cr atoms can significantly influence the magnetic spin structures in $Cr_{1+x}Te_2$ systems. Such compositional sensitivity renders $Cr_{1+x}Te_2$ an excellent and tunable platform for exploring a variety of exotic spin structures, including skyrmions, antiskyrmions, and complex non-coplanar structures etc. Our findings, in tandem with previous reports, underscore the versatility of this material system for investigating the interplay between composition, magnetic anisotropy, and topological spin structures.

## 3. Conclusion

In summary, we have systematically investigated the magnetic and transport properties of $Cr_{1.61}Te_2$ single crystals. $Cr_{1.61}Te_2$ exhibits a stable ferromagnetic behavior at ambient condition with $T_C \sim 326$ K. Magnetic properties study reveals an in-plane magnetic anisotropy in this compound, and $K_U$ displays non-monotonic temperature dependence, reaching a maximum value of $3.73 \times 10^5$ J/m$^3$ near $T^{\#}$, which may be useful in the information technology. Moreover, the MR and Hall effect exhibit distinct behaviors in different temperature ranges, indicating the evolution of spin structures in $Cr_{1.61}Te_2$ with decreasing temperature. When $T^* > T > T^{\#}$, a large THE, comparable to AHE and induced by microscopic non-coplanar spin structures, is observed. Furthermore, another topological spin structures may emerge at low temperatures when $T < T^{\#}$. These findings indicate $Cr_{1.61}Te_2$ could serve as an excellent platform for investigating the evolution of spin structures with only temperature, without invoking other tuning parameters like pressure or ionic liquid gating. Potential applications of these anisotropic magnetic properties and topological spin structure revealed in this study as information carriers can thus be envisaged.

## 4. Experimental Section

*Crystal Growth:* Single crystals of $Cr_{1.61}Te_2$ were synthesized by using the chemical vapor transport (CVT) method. High-purity powders of chromium (Cr, 99.99%, Alfa Aesar) and tellurium (Te, 99.99%, Alfa Aesar) were thoroughly mixed in an argon-filled glove box, employing iodine as the transport agent. The mixture was then sealed in an evacuated quartz tube and placed in a tubular furnace. The quartz tube was gradually heated to 1000 °C in the source zone and to 900 °C in the deposition zone at a rate of 1 °C/min and maintained at these temperatures for 14 days. Subsequently, the furnace was cooled to room temperature. Bright single crystals were collected from the cooler side of the tube.

*Sample Characterizations:* The chemical compositions were characterized by EDX (HGSTFlexSEM-1000). XRD was conducted by using the x-ray diffractometer (Bruker D2 PHASER). The specimens for the scanning transmission electron microscopy (STEM) measurements were prepared via the focused ion beam (FIB) milling technique (Helios Nanolab 600i; FEI). Atomic-scale HAADF-STEM images were performed on a FEI Titan$^3$ Themis G3 60–300.

*Electrical Transport Measurements:* Gold wires were used as electrodes of the bulk devices, and they were connected by silver epoxy. (Magneto-)resistance and the Hall effect were measured using the standard four-probe method in applied fields up to 9 T in the Cryogenic system. By reversing the field polarities, the even signal in field was defined as MR and the odd component was calculated as the Hall resistivity.

*Magnetization Measurements:* Magnetization measurements were conducted using a superconducting quantum interference device magnetometer (MPMS, Quantum Design). Both zero-field-cooled and field-cooled modes were employed.

**Acknowledgements**

Y. Huang and N. Zuo contributed equally to this work. The authors would like to acknowledge M. Smidman, N. E. Hussey, A. F. Bangura and J. H. Dai for stimulating discussions. This work was supported by National Natural Science Foundation of China (Grants No. 12274369, No. 12304071, No.12204265) and the Young Scientists of Taishan Scholarship (Grants No. tsqn202408168 and No. tsqn202211128). A portion of work was supported by Zhejiang Provincial Natural Science Foundation of China

(Grants No. LZ25A040003 and No. LZ23A040002). D. Q. was financially supported by the National Key Projects for Research & Development of China (Grant No. 2022YFA1402400).

**Data Availability Statement**

The data that support the findings of this study are available from the corresponding author upon reasonable request.

**References**

[1] Li H, Ruan S and Zeng Y-J. Intrinsic Van Der Waals Magnetic Materials from Bulk to the 2D Limit: New Frontiers of Spintronics. Advanced Materials, 2019, 31(27): 1900065.

[2] Chen X, Shao Y-T, Chen R, Susarla S, Hogan T, He Y, Zhang H, Wang S, Yao J, Ercius P, Muller D A, Ramesh R and Birgeneau R J. Pervasive beyond Room-Temperature Ferromagnetism in a Doped van der Waals Magnet. Physical Review Letters, 2022, 128(21): 217203.

[3] Yang S, Zhang T and Jiang C. van der Waals Magnets: Material Family, Detection and Modulation of Magnetism, and Perspective in Spintronics. Advanced Science, 2021, 8(2): 2002488.

[4] Lv X, Huang Y, Pei K, Yang C, Zhang T, Li W, Cao G, Zhang J, Lai Y and Che R. Manipulating the Magnetic Bubbles and Topological Hall Effect in 2D Magnet $Fe_5GeTe_2$. Advanced Functional Materials, 2024, 34(11): 2308560.

[5] Deng Y, Yu Y, Song Y, Zhang J, Wang N Z, Sun Z, Yi Y, Wu Y Z, Wu S, Zhu J, Wang J, Chen X H and Zhang Y. Gate-tunable room-temperature ferromagnetism in two-dimensional $Fe_3GeTe_2$. Nature, 2018, 563(7729): 94-9.

[6] Wang H, Lu H, Guo Z, Li A, Wu P, Li J, Xie W, Sun Z, Li P, Damas H, Friedel A M, Migot S, Ghanbaja J, Moreau L, Fagot-Revurat Y, Petit-Watelot S, Hauet T, Robertson J, Mangin S, Zhao W and Nie T. Interfacial engineering of ferromagnetism in wafer-scale van der Waals $Fe_4GeTe_2$ far above room temperature. Nature Communications, 2023, 14(1): 2483.

[7] Lv X, Lv H, Huang Y, Zhang R, Qin G, Dong Y, Liu M, Pei K, Cao G, Zhang J, Lai Y and Che R. Distinct skyrmion phases at room temperature in two-dimensional ferromagnet $Fe_3GaTe_2$. Nature Communications, 2024, 15(1): 3278.

[8] Zhang G, Guo F, Wu H, Wen X, Yang L, Jin W, Zhang W and Chang H. Above-room-temperature strong intrinsic ferromagnetism in 2D van der Waals $Fe_3GaTe_2$ with large perpendicular magnetic anisotropy. Nature Communications, 2022, 13(1): 5067.

[9] Gish J T, Lebedev D, Song T W, Sangwan V K and Hersam M C. Van der Waals opto-spintronics. Nature Electronics, 2024, 7(5): 336-47.

[10] Liang S-J, Cheng B, Cui X and Miao F. Van der Waals Heterostructures for High-Performance Device Applications: Challenges and Opportunities. Advanced Materials, 2020, 32(27): 1903800.

[11] Gong C, Li L, Li Z, Ji H, Stern A, Xia Y, Cao T, Bao W, Wang C, Wang Y, Qiu Z Q, Cava R J, Louie S G, Xia J and Zhang X. Discovery of intrinsic ferromagnetism in two-dimensional van der Waals crystals. Nature, 2017, 546(7657): 265-9.

[12] Wu H, Zhang W, Yang L, Wang J, Li J, Li L, Gao Y, Zhang L, Du J, Shu H and Chang H. Strong intrinsic room-temperature ferromagnetism in freestanding non-van der Waals ultrathin 2D crystals. Nature Communications, 2021, 12(1): 5688.

[13] He K, Bian M, Seddon S D, Jagadish K, Mucchietto A, Ren H, Kirstein E, Asadi R, Bai J, Yao C, Pan S, Yu J-X, Milde P, Huai C, Hui H, Zang J, Sabirianov R, Cheng X M, Miao G, Xing H, Shao Y-T, Crooker S A, Eng L, Hou Y, Bird J P and Zeng H. Unconventional Anomalous Hall Effect Driven by Self-Intercalation in Covalent 2D Magnet $Cr_2Te_3$. Advanced Science, 2025, 12(2): 2407625.

[14] Peng J, Su Y, Lv H, Wu J, Liu Y, Wang M, Zhao J, Guo Y, Wu X, Wu C and Xie Y. Even–Odd-Layer-Dependent Ferromagnetism in 2D Non-van-der-Waals $CrCuSe_2$. Advanced Materials, 2023, 35(16): 2209365.

[15] Bian M, N. K A, Mengjiao H, Wenjie L, Sichen W, Xuezeng T, B. E D, Fan S, Keke H, Haolei H, Fei Y, Renat S, P. B J, Chunlei Y, Jianwei M, Junhao L, A. C S, Yanglong H and And Zeng H. Covalent 2D $Cr_2Te_3$ ferromagnet. Materials Research Letters, 2021, 9(5): 205-12.

[16] Ni Y, Wang T, Wang J, Guo Y, Huang T, Qiao X, Zhang W, Zhang Z, Chen X, Li T and Min T. Evolution of Microscopic Magnetic Domains in Quasi-2D $Cr_{0.92}Te$ at Room Temperature. Advanced Physics Research, 2024, 3(5): 2300116.

[17] Chua R, Zhou J, Yu X, Yu W, Gou J, Zhu R, Zhang L, Liu M, Breese M B H, Chen W, Loh K P, Feng Y P, Yang M, Huang Y L and Wee A T S. Room Temperature Ferromagnetism of Monolayer Chromium Telluride with Perpendicular Magnetic Anisotropy. Advanced Materials, 2021, 33(42): 2103360.

[18] Fu L, Llacsahuanga Allcca A E and Chen Y P. Coexisting Ferromagnetic–Antiferromagnetic State and Giant Anomalous Hall Effect in Chemical Vapor Deposition-Grown 2D $Cr_5Te_8$. ACS Nano, 2024, 18(49): 33381-9.

[19] Dijkstra J, Weitering H H, Bruggen C F V, Haas C and Groot R a D. Band-structure calculations, and magnetic and transport properties of ferromagnetic chromium tellurides ($CrTe$, $Cr_3Te_4$, $Cr_2Te_3$). Journal of Physics: Condensed Matter, 1989, 1(46): 9141.

[20] Ipser H, Komarek K L and Klepp K O. Transition metal-chalcogen systems viii: The Cr Te phase diagram. Journal of the Less Common Metals, 1983, 92(2): 265-82.

[21] Meng L, Zhou Z, Xu M, Yang S, Si K, Liu L, Wang X, Jiang H, Li B, Qin P, Zhang P, Wang J, Liu Z, Tang P, Ye Y, Zhou W, Bao L, Gao H-J and Gong Y. Anomalous thickness dependence of Curie temperature in air-stable two-dimensional ferromagnetic 1T-CrTe2 grown by chemical vapor deposition. Nature Communications, 2021, 12(1): 809.

[22] Zhang X, Lu Q, Liu W, Niu W, Sun J, Cook J, Vaninger M, Miceli P F, Singh D J, Lian S-W, Chang T-R, He X, Du J, He L, Zhang R, Bian G and Xu Y. Room-temperature intrinsic ferromagnetism in epitaxial CrTe2 ultrathin films. Nature Communications, 2021, 12(1): 2492.

[23] Zhang X, Li Y, Lu Q, Xiang X, Sun X, Tang C, Mahdi M, Conner C, Cook J, Xiong Y, Inman J, Jin W, Liu C, Cai P, Santos E J G, Phatak C, Zhang W, Gao N, Niu W, Bian G, Li P, Yu D and Long S. Epitaxial Growth of Large-Scale 2D CrTe2 Films on Amorphous Silicon Wafers With Low Thermal Budget. Advanced Materials, 2024, 36(24): 2311591.

[24] Pandey S and Mukhopadhyay S. Competing magnetocrystalline anisotropy and spin-dimension crossover in $Cr_{1+x}Te_2$. Physical Review B, 2024, 110(6): L060405.

[25]     Zhang L-Z, Zhang A-L, He X-D, Ben X-W, Xiao Q-L, Lu W-L, Chen F, Feng Z, Cao S, Zhang J and Ge J-Y. Critical behavior and magnetocaloric effect of the quasi-two-dimensional room-temperature ferromagnet $Cr_4Te_5$. Physical Review B, 2020, 101(21): 214413.

[26]     Zhang C, Liu C, Zhang J, Yuan Y, Wen Y, Li Y, Zheng D, Zhang Q, Hou Z, Yin G, Liu K, Peng Y and Zhang X-X. Room-Temperature Magnetic Skyrmions and Large Topological Hall Effect in Chromium Telluride Engineered by Self-Intercalation. Advanced Materials, 2023, 35(1): 2205967.

[27]     Liu Y, Liu Y, Dan J, Liu W, Wang L, Hu K, Wang W, Zhang L, Ge B, Du H and Song D. Atomic-Scale Order and Disorder Induced Diverse Topological Spin Textures in Self-Intercalated Van der Waals Magnets $Cr_{1+\delta}Te_2$. Advanced Functional Materials, 2025, 35(5): 2414699.

[28]     Yang J, Zhu C, Deng Y, Tang B and Liu Z. Magnetism of two-dimensional chromium tellurides. iScience, 2023, 26(5): 106567.

[29]     Lasek K, Coelho P M, Gargiani P, Valvidares M, Mohseni K, Meyerheim H L, Kostanovskiy I, Zberecki K and Batzill M. Van der Waals epitaxy growth of 2D ferromagnetic $Cr_{(1+\delta)}Te_2$ nanolayers with concentration-tunable magnetic anisotropy. Applied Physics Reviews, 2022, 9(1): 011409.

[30]     Li B, Deng X, Shu W, Cheng X, Qian Q, Wan Z, Zhao B, Shen X, Wu R, Shi S, Zhang H, Zhang Z, Yang X, Zhang J, Zhong M, Xia Q, Li J, Liu Y, Liao L, Ye Y, Dai L, Peng Y, Li B and Duan X. Air-stable ultrathin $Cr_3Te_4$ nanosheets with thickness-dependent magnetic biskyrmions. Materials Today, 2022, 57: 66-74.

[31]     Huang M, Gao L, Zhang Y, Lei X, Hu G, Xiang J, Zeng H, Fu X, Zhang Z, Chai G, Peng Y, Lu Y, Du H, Chen G, Zang J and Xiang B. Possible Topological Hall Effect above Room Temperature in Layered $Cr_{1.2}Te_2$ Ferromagnet. Nano Letters, 2021, 21(10): 4280-6.

[32]     Purwar S, Low A, Bose A, Narayan A and Thirupathaiah S. Investigation of the anomalous and topological Hall effects in layered monoclinic ferromagnet $Cr_{2.76}Te_4$. Physical Review Materials, 2023, 7(9): 094204.

[33]     Liu J, Ding B, Liang J, Li X, Yao Y and Wang W. Magnetic Skyrmionic Bubbles at Room Temperature and Sign Reversal of the Topological Hall Effect in a Layered Ferromagnet $Cr_{0.87}Te$. ACS Nano, 2022, 16(9): 13911-8.

[34]     Purwar S, Changdar S, Ghosh S, Bhowmik T K and Thirupathaiah S. Intricate magnetic interactions and topological Hall effect observed in itinerant room-temperature layered ferromagnet $Cr_{0.83}Te$. Acta Materialia, 2024, 271: 119898.

[35]     Wang A, Rahman A, Du Z, Zhao J, Meng F, Liu W, Fan J, Ma C, Ge M, Pi L, Zhang L and Zhang Y. Field-dependent anisotropic room-temperature ferromagnetism in $Cr_3Te_4$. Physical Review B, 2023, 108(9): 094429.

[36]     Liu Y and Petrovic C. Critical behavior of the quasi-two-dimensional weak itinerant ferromagnet trigonal chromium telluride $Cr_{0.62}Te$. Physical Review B, 2017, 96(13): 134410.

[37]     Li C, Liu K, Jiang D, Jin C, Pei T, Wen T, Yue B and Wang Y. Diverse thermal expansion behaviors in ferromagnetic $Cr_{1-\delta}Te$ with NiAs-type, defective structures. Inorganic Chemistry, 2022, 61(37): 14641-7.

[38]     Bertaut E, Roult G, Aleonard R, Pauthenet R, Chevreton M and Jansen R. Structures

magnétiques de Cr$_3$X$_4$ (X= S, Se, Te). Journal de Physique, 1964, 25(5): 582-95.

[39] Andresen A F, Zeppezauer E, Boive T, Nordström B and Brändén C. Magnetic structure of Cr$_2$Te$_3$, Cr$_3$te$_4$, and Cr$_5$Te$_6$. Acta Chem Scand, 1970, 24(10): 3495-509.

[40] Yamaguchi M and Hashimoto T. Magnetic Properties of Cr3Te4 in Ferromagnetic Region. Journal of the Physical Society of Japan, 1972, 32(3): 635-8.

[41] Du-Xing C, Pardo E and Sanchez A. Demagnetizing factors of rectangular prisms and ellipsoids. IEEE Transactions on Magnetics, 2002, 38(4): 1742-52.

[42] Sucksmith W and Thompson J E. The magnetic anisotropy of cobalt. Proceedings of the Royal Society of London Series A Mathematical and Physical Sciences, 1954, 225(1162): 362-75.

[43] Johnson M, Bloemen P, Den Broeder F and De Vries J. Magnetic anisotropy in metallic multilayers. Reports on Progress in Physics, 1996, 59(11): 1409.

[44] Hou Z, Ren W, Ding B, Xu G, Wang Y, Yang B, Zhang Q, Zhang Y, Liu E, Xu F, Wang W, Wu G, Zhang X, Shen B and Zhang Z. Observation of Various and Spontaneous Magnetic Skyrmionic Bubbles at Room Temperature in a Frustrated Kagome Magnet with Uniaxial Magnetic Anisotropy. Advanced Materials, 2017, 29(29): 1701144.

[45] Ueda K and Moriya T. Contribution of Spin Fluctuations to the Electrical and Thermal Resistivities of Weakly and Nearly Ferromagnetic Metals. Journal of the Physical Society of Japan, 1975, 39(3): 605-15.

[46] Raquet B, Viret M, Sondergard E, Cespedes O and Mamy R. Electron-magnon scattering and magnetic resistivity in 3$d$ ferromagnets. Physical Review B, 2002, 66(2): 024433.

[47] Hwang H Y, Cheong S W, Ong N P and Batlogg B. Spin-Polarized Intergrain Tunneling in La$_{2/3}$Sr$_{1/3}$MnO$_3$. Physical Review Letters, 1996, 77(10): 2041-4.

[48] Telford E J, Dismukes A H, Dudley R L, Wiscons R A, Lee K, Chica D G, Ziebel M E, Han M-G, Yu J, Shabani S, Scheie A, Watanabe K, Taniguchi T, Xiao D, Zhu Y, Pasupathy A N, Nuckolls C, Zhu X, Dean C R and Roy X. Coupling between magnetic order and charge transport in a two-dimensional magnetic semiconductor. Nature Materials, 2022, 21(7): 754-60.

[49] Lin Z, Lohmann M, Ali Z A, Tang C, Li J, Xing W, Zhong J, Jia S, Han W, Coh S, Beyermann W and Shi J. Pressure-induced spin reorientation transition in layered ferromagnetic insulator Cr$_2$Ge$_2$Te$_6$. Physical Review Materials, 2018, 2(5): 051004.

[50] Deng Y, Xiang Z, Lei B, Zhu K, Mu H, Zhuo W, Hua X, Wang M, Wang Z, Wang G, Tian M and Chen X. Layer-Number-Dependent Magnetism and Anomalous Hall Effect in van der Waals Ferromagnet Fe$_5$GeTe$_2$. Nano Letters, 2022, 22(24): 9839-46.

[51] Kanazawa N, Onose Y, Arima T, Okuyama D, Ohoyama K, Wakimoto S, Kakurai K, Ishiwata S and Tokura Y. Large Topological Hall Effect in a Short-Period Helimagnet MnGe. Physical Review Letters, 2011, 106(15): 156603.

[52] He Y, Kroder J, Gayles J, Fu C, Pan Y, Schnelle W, Felser C and Fecher G H. Large topological Hall effect in an easy-cone ferromagnet (Cr$_{0.9}$B$_{0.1}$)Te. Applied Physics Letters, 2020, 117(5).

[53] Neubauer A, Pfleiderer C, Binz B, Rosch A, Ritz R, Niklowitz P G and Böni P. Topological Hall Effect in the *A* Phase of MnSi. Physical Review Letters, 2009, 102(18): 186602.

[54] Taguchi Y, Oohara Y, Yoshizawa H, Nagaosa N and Tokura Y. Spin Chirality, Berry Phase, and Anomalous Hall Effect in a Frustrated Ferromagnet. Science, 2001, 291(5513): 2573-

6.

[55] Schulz T, Ritz R, Bauer A, Halder M, Wagner M, Franz C, Pfleiderer C, Everschor K, Garst M and Rosch A. Emergent electrodynamics of skyrmions in a chiral magnet. Nature Physics, 2012, 8(4): 301-4.

[56] Zhang X, Ambhire S C, Lu Q, Niu W, Cook J, Jiang J S, Hong D, Alahmed L, He L, Zhang R, Xu Y, Zhang S S L, Li P and Bian G. Giant Topological Hall Effect in van der Waals Heterostructures of $CrTe_2/Bi_2Te_3$. ACS Nano, 2021, 15(10): 15710-9.

[57] Gallagher J C, Meng K Y, Brangham J T, Wang H L, Esser B D, Mccomb D W and Yang F Y. Robust Zero-Field Skyrmion Formation in FeGe Epitaxial Thin Films. Physical Review Letters, 2017, 118(2): 027201.

[58] Maryenko D, Mishchenko A S, Bahramy M S, Ernst A, Falson J, Kozuka Y, Tsukazaki A, Nagaosa N and Kawasaki M. Observation of anomalous Hall effect in a non-magnetic two-dimensional electron system. Nature Communications, 2017, 8(1): 14777.

[59] Onoda S, Sugimoto N and Nagaosa N. Quantum transport theory of anomalous electric, thermoelectric, and thermal Hall effects in ferromagnets. Physical Review B, 2008, 77(16): 165103.

[60] Li Y, Kanazawa N, Yu X Z, Tsukazaki A, Kawasaki M, Ichikawa M, Jin X F, Kagawa F and Tokura Y. Robust Formation of Skyrmions and Topological Hall Effect Anomaly in Epitaxial Thin Films of MnSi. Physical Review Letters, 2013, 110(11): 117202.

[61] Leroux M, Stolt M J, Jin S, Pete D V, Reichhardt C and Maiorov B. Skyrmion Lattice Topological Hall Effect near Room Temperature. Scientific Reports, 2018, 8(1): 15510.

[62] Pahari R, Balasubramanian B, Ullah A, Manchanda P, Komuro H, Streubel R, Klewe C, Valloppilly S R, Shafer P, Dev P, Skomski R and Sellmyer D J. Peripheral chiral spin textures and topological Hall effect in CoSi nanomagnets. Physical Review Materials, 2021, 5(12): 124418.

[63] Li Y, Xu G, Ding B, Liu E, Wang W and Liu Z. The electronic and magnetic properties and topological Hall effect in hexagonal MnNiGa alloy films by varying Mn contents. Journal of Alloys and Compounds, 2017, 725: 1324-9.

[64] He Y, Schneider S, Helm T, Gayles J, Wolf D, Soldatov I, Borrmann H, Schnelle W, Schaefer R, Fecher G H, Rellinghaus B and Felser C. Topological Hall effect arising from the mesoscopic and microscopic non-coplanar magnetic structure in MnBi. Acta Materialia, 2022, 226: 117619.

[65] Vir P, Gayles J, Sukhanov A S, Kumar N, Damay F, Sun Y, Kübler J, Shekhar C and Felser C. Anisotropic topological Hall effect with real and momentum space Berry curvature in the antiskrymion-hosting Heusler compound $Mn_{1.4}PtSn$. Physical Review B, 2019, 99(14): 140406.

[66] Gong G, Xu L, Bai Y, Wang Y, Yuan S, Liu Y and Tian Z. Large topological Hall effect near room temperature in noncollinear ferromagnet $LaMn_2Ge_2$ single crystal. Physical Review Materials, 2021, 5(3): 034405.

[67] Dhakal G, Cheenicode Kabeer F, Pathak A K, Kabir F, Poudel N, Filippone R, Casey J, Pradhan Sakhya A, Regmi S, Sims C, Dimitri K, Manfrinetti P, Gofryk K, Oppeneer P M and Neupane M. Anisotropically large anomalous and topological Hall effect in a kagome magnet. Physical Review B, 2021, 104(16): L161115.

[68] Li H, Ding B, Chen J, Li Z, Liu E, Xi X, Wu G and Wang W. Large anisotropic topological

Hall effect in a hexagonal non-collinear magnet $Fe_5Sn_3$. Applied Physics Letters, 2020, 116(18).

[69]  Casas B W, Li Y, Moon A, Xin Y, Mckeever C, Macy J, Petford-Long A K, Phatak C M, Santos E J G, Choi E S and Balicas L. Coexistence of Merons with Skyrmions in the Centrosymmetric Van Der Waals Ferromagnet $Fe_{5-x}GeTe_2$. Advanced Materials, 2023, 35(17): 2212087.

[70]  Gao Y, Yin Q, Wang Q, Li Z, Cai J, Zhao T, Lei H, Wang S, Zhang Y and Shen B. Spontaneous (Anti)meron Chains in the Domain Walls of van der Waals Ferromagnetic $Fe_{5-x}GeTe_2$. Advanced Materials, 2020, 32(48): 2005228.

[71]  Wang Y, Yan J, Li J, Wang S, Song M, Song J, Li Z, Chen K, Qin Y, Ling L, Du H, Cao L, Luo X, Xiong Y and Sun Y. Magnetic anisotropy and topological Hall effect in the trigonal chromium tellurides $Cr_5Te_8$. Physical Review B, 2019, 100(2): 024434.

[72]  Liu E, Sun Y, Kumar N, Muechler L, Sun A, Jiao L, Yang S-Y, Liu D, Liang A, Xu Q, Kroder J, Süß V, Borrmann H, Shekhar C, Wang Z, Xi C, Wang W, Schnelle W, Wirth S, Chen Y, Goennenwein S T B and Felser C. Giant anomalous Hall effect in a ferromagnetic kagome-lattice semimetal. Nature Physics, 2018, 14(11): 1125-31.

[73]  Kim K, Seo J, Lee E, Ko K T, Kim B S, Jang B G, Ok J M, Lee J, Jo Y J, Kang W, Shim J H, Kim C, Yeom H W, Il Min B, Yang B-J and Kim J S. Large anomalous Hall current induced by topological nodal lines in a ferromagnetic van der Waals semimetal. Nature Materials, 2018, 17(9): 794-9.

[74]  Iguchi S, Hanasaki N and Tokura Y. Scaling of anomalous hall resistivity in $Nd_2(Mo_{(1-x)}Nb_{(x)})_2O_7$ with spin chirality. Physical Review Letters, 2007, 99(7): 077202.

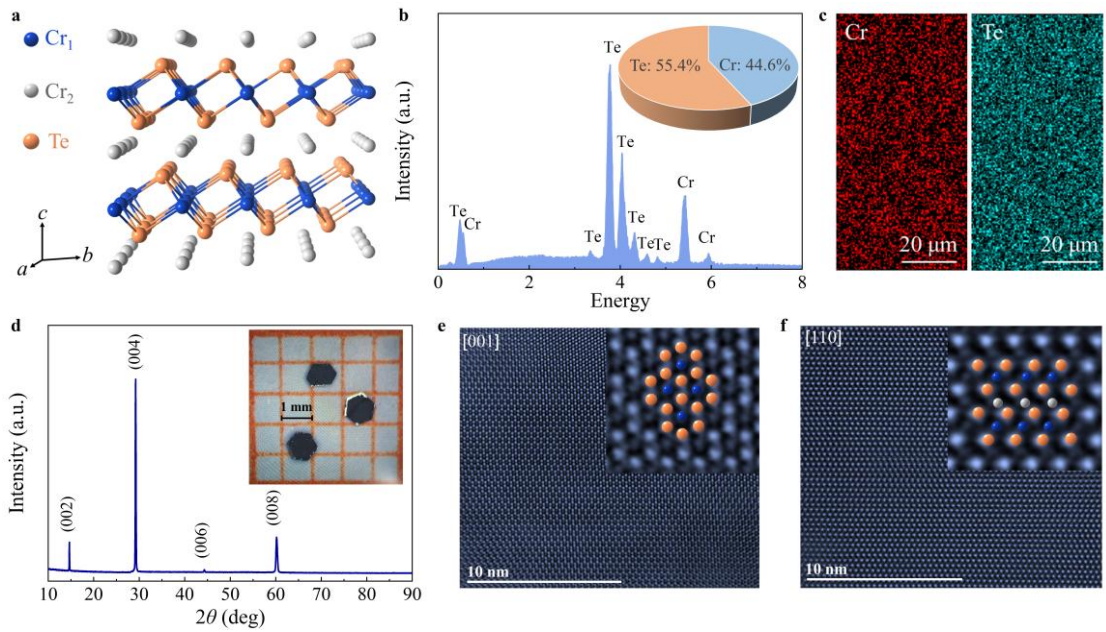

Figure 1 **a** Crystal structure of $Cr_{1+x}Te_2$, where $Cr_1$ and $Cr_2$ denote atoms within and between the $CrTe_2$ layers, respectively. **b** A typical EDX pattern of $Cr_{1.61}Te_2$. The inset displays the element molar ratio. **c** The EDX mapping of Te and Cr for the as-grown single crystal revealing the spatial distribution of the constituent elements. **d** XRD patterns of $Cr_{1.61}Te_2$, showing only the (00l) Bragg peaks, which indicate that the exposed surface corresponds to the crystalline *ab* plane. The inset presents an optical image of the as-grown single crystal, which exhibits a hexagonal morphology consistent with the expected crystal structure. **e**, **f** HAADF-STEM images taken along the [001] and [1$\bar{1}$0] axes of $Cr_{1.61}Te_2$ single crystals, respectively. The scale bars are 10 nm. The top insets show the enlarged image of the alternate arrangement of Te (orange particles), Cr1 (blue particles), and Cr2 (gray particles) atoms.

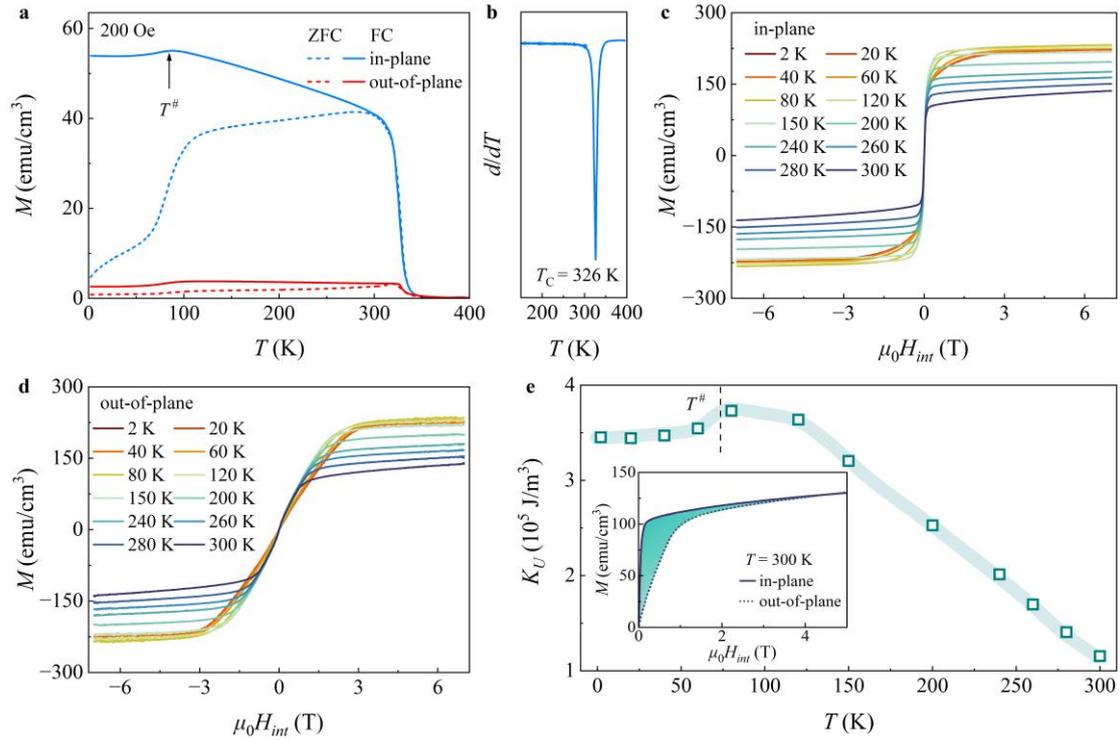

Figure 2 **a** Temperature dependence of magnetization in $Cr_{1.61}Te_2$ under a magnetic field of 100 Oe in ZFC and FC modes, measured for both in-plane and out-of-plane field configurations. **b** The derivative of the *M-T* curve under the out-of-plane field in the FC mode. **c**, **d** Field dependent magnetization measured at different temperatures for in-plane and out-of-plane field configurations, respectively. **e** Temperature dependence of $K_U$, with a trendline provided as a guide to the eye. The inset displays the in-plane and out-of-plane *M-H* curves measured at 300 K. The shaded area represents the magnetic anisotropy energy density $K_U = \mu_0 \int_0^{M_s}[H_{ab}(M) - H_c(M)]dM$.

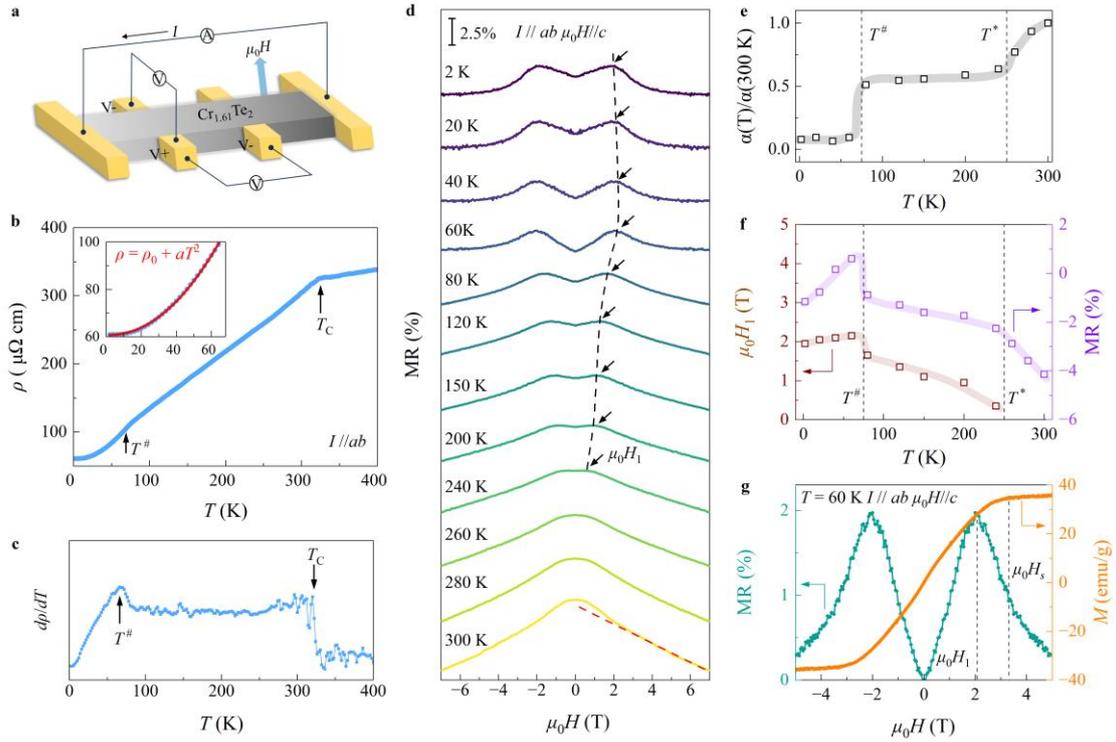

Figure 3 **a** Schematic illustration of the $Cr_{1.61}Te_2$ Hall device. **b** Temperature-dependent in-plane resistivity of the $Cr_{1.61}Te_2$ single crystal. The inset shows the low-temperature resistivity fitted by $\rho(T) = \rho_0 + T^2$. **c** The derivative of the $\rho$-$T$ curve. **d** In-plane MR at different temperatures under the out-of-plane magnetic field for $Cr_{1.61}Te_2$. The red dashed line is a linear guiding line. $\mu_0 H_1$ denotes the peak in the MR. **e** Normalized slope magnitude $\alpha(T)/\alpha(300\ K)$ of LNMR plotted against temperature from 2 K to 300 K, with trendlines provided as a guide to the eye. **f** Temperature dependence of $\mu_0 H_1$ and MR at 7 T, with trendlines provided as a guide to the eye. **g** Left column: In-plane MR at 60 K with the out-of-plane magnetic field. Right column: *M-H* curve at 60 K with the out-of-plane magnetic field.

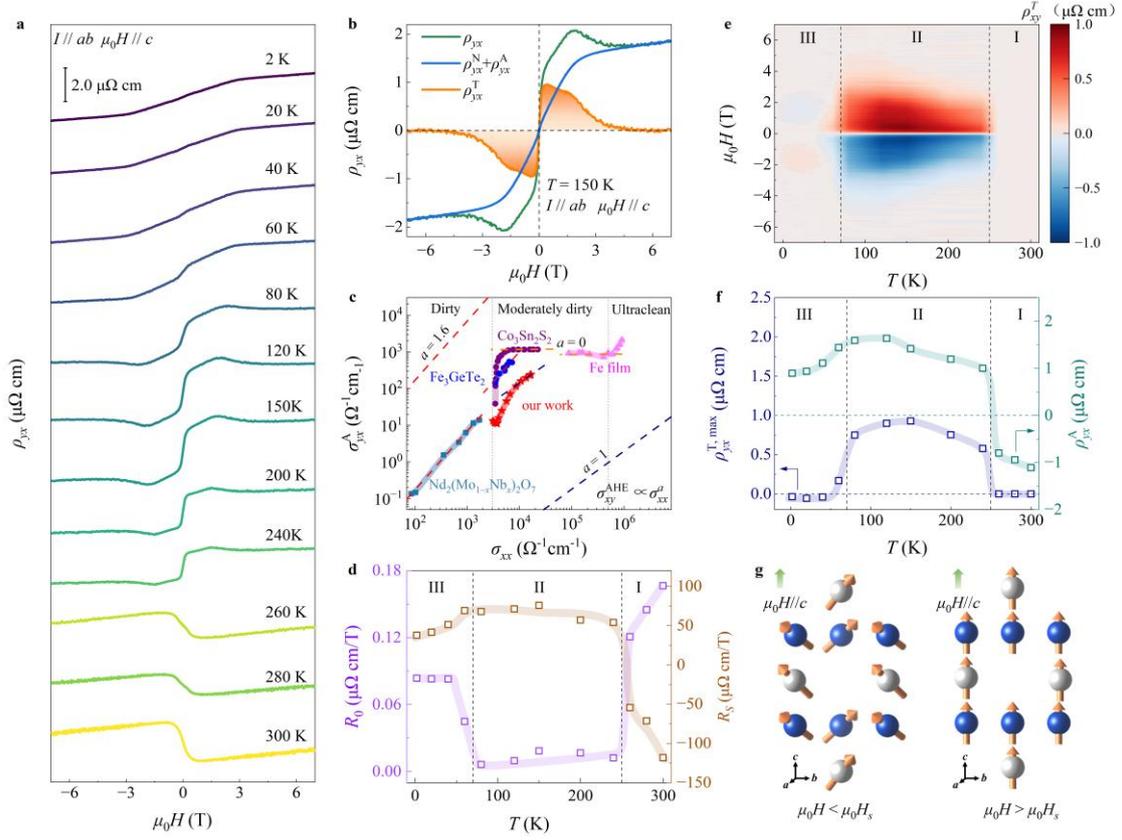

Figure 4 **a** Magnetic-field dependence of Hall resistivity at various temperatures for $Cr_{1.61}Te_2$. **b** Parsing the different contributions to the Hall effect at 150 K, which is the sum of the normal Hall resistivity $\rho_{yx}^N$, anomalous Hall resistivity $\rho_{yx}^A$ and the topological Hall resistivity $\rho_{yx}^T$. **c** Anomalous Hall conductivity $\sigma_{yx}^A$ versus longitudinal conductivity $\sigma_{xx}$ for $Cr_{1.61}Te_2$, along with the data for other ferromagnets on a log−log scale (including Refs. [59, 72-74]). **d** Temperature dependence of normal Hall coefficient $R_0$ and anomalous Hall coefficient $R_S$. **e** Contour plot of the as-obtained topological Hall resistivity $\rho_{yx}^T$. **f** Temperature dependence of the maximum magnitude of topological Hall resistivity $\rho_{yx}^{T,max}$ and anomalous Hall resistivity $\rho_{yx}^A$. Negative symbol represents a sign reversal for THE and AHE. **g** Schematic illustration of the Cr spin configurations for $\mu_0 H < \mu_0 H_s$ (left) and for $\mu_0 H > \mu_0 H_s$ (right).

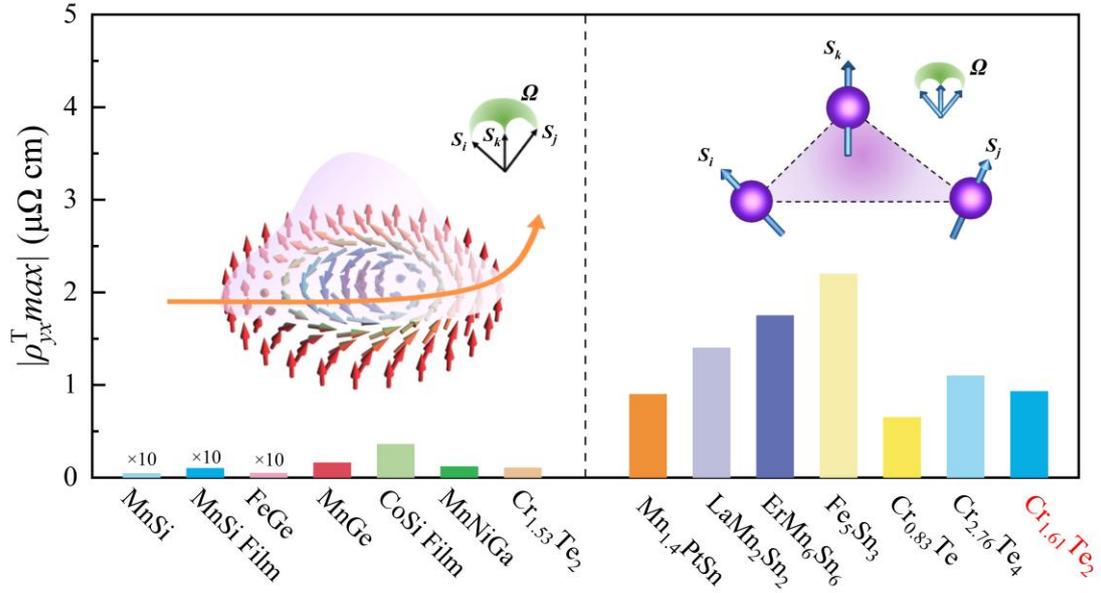

Figure 5. Comparison of the $\rho_{yx}^{T,max}$ values observed in various materials exhibiting mesoscopic skyrmions and microscopic non-coplanar spin structures. The data were compiled from published literature (including Refs. [26, 32, 34, 51, 53, 60-63, 65-68]). Overall, the THE induced by skyrmions is relatively weak. Moreover, the length scale of spin textures for skyrmions (ranging from 1 to $10^3$ nm) is significantly larger than that of non-coplanar spin structures (ranging from 0.1 to 1 nm). The left and right insets depict the schematic representations of THE induced by skyrmions and non-coplanar spin structures, respectively.

Table 1. Summary of some key parameters, including temperature regimes, magnetic anisotropy, THE magnitude, and proposed spin textures, in $Cr_{1+x}Te_2$ systems. Out-of-plane magnetic anisotropy is abbreviated to OOP, while IP denotes in-plane magnetic anisotropy.

| Compound | Temperature Ranges | Magnetic Anisotropy | Max THE magnitude | Proposed Spin Structures |
|---|---|---|---|---|
| $Cr_{1.25}Te_2$[71] | 160-2 K | OOP | 0.18 µΩ cm | noncoplanar spin structure |
| $Cr_{1.74}Te_2$[33] | 350-10 K | OOP | 0.20 µΩ cm | skyrmions |
| $Cr_{1.38}Te_2$[32] | 250-3 K | IP | 1.1 µΩ cm | noncoplanar spin structure |
| $Cr_{1.66}Te_2$[34] | 240-10 K | IP | 0.62 µΩ cm | non-coplanar spin structure |
| $Cr_{1.53}Te_2$[26] | 295-250 K | OOP | 106 nΩ cm | skyrmions |
| | 250-90 K | OOP | 75 nΩ cm | non-collinear FM |
| | 90-2 K | OOP | 175 nΩ cm | antiskyrmions |
| $Cr_{1.61}Te_2$ (this work) | 300-260 K | IP | ~ 0 | — |
| | 260-75 K | IP | 0.93 µΩ cm | non-coplanar spin structure |
| | 75-2 K | IP | 56 nΩ cm | merons (antimerons) or non-coplanar spin structures |